\input amssym.def
\input epsf


\magnification=\magstephalf
\hsize=14.0 true cm
\vsize=19 true cm
\hoffset=1.0 true cm
\voffset=2.0 true cm

\abovedisplayskip=12pt plus 3pt minus 3pt
\belowdisplayskip=12pt plus 3pt minus 3pt
\parindent=1em


\font\sixrm=cmr6
\font\eightrm=cmr8
\font\ninerm=cmr9

\font\sixi=cmmi6
\font\eighti=cmmi8
\font\ninei=cmmi9

\font\sixsy=cmsy6
\font\eightsy=cmsy8
\font\ninesy=cmsy9

\font\sixbf=cmbx6
\font\eightbf=cmbx8
\font\ninebf=cmbx9

\font\eightit=cmti8
\font\nineit=cmti9

\font\eightsl=cmsl8
\font\ninesl=cmsl9

\font\sixss=cmss8 at 8 true pt
\font\sevenss=cmss9 at 9 true pt
\font\eightss=cmss8
\font\niness=cmss9
\font\tenss=cmss10

 at 12 true pt
\font\bigrm=cmr10 at 12 true pt
\font\bigbf=cmbx10 at 12 true pt

 at 16 true pt
 at 16 true pt
 at 16 true pt

\catcode`@=11
\newfam\ssfam

\def\tenpoint{\def\rm{\fam0\tenrm}%
    \textfont0=\tenrm \scriptfont0=\sevenrm \scriptscriptfont0=\fiverm
    \textfont1=\teni  \scriptfont1=\seveni  \scriptscriptfont1=\fivei
    \textfont2=\tensy \scriptfont2=\sevensy \scriptscriptfont2=\fivesy
    \textfont3=\tenex \scriptfont3=\tenex   \scriptscriptfont3=\tenex
    \textfont\itfam=\tenit                  \def\it{\fam\itfam\tenit}%
    \textfont\slfam=\tensl                  \def\sl{\fam\slfam\tensl}%
    \textfont\bffam=\tenbf \scriptfont\bffam=\sevenbf
    \scriptscriptfont\bffam=\fivebf
                                            \def\bf{\fam\bffam\tenbf}%
    \textfont\ssfam=\tenss \scriptfont\ssfam=\sevenss
    \scriptscriptfont\ssfam=\sevenss
                                            \def\ss{\fam\ssfam\tenss}%
    \normalbaselineskip=13pt
    \setbox\strutbox=\hbox{\vrule height8.5pt depth3.5pt width0pt}%
    \let\big=\tenbig
    \normalbaselines\rm}

\def\ninepoint{\def\rm{\fam0\ninerm}%
    \textfont0=\ninerm      \scriptfont0=\sixrm
                            \scriptscriptfont0=\fiverm
    \textfont1=\ninei       \scriptfont1=\sixi
                            \scriptscriptfont1=\fivei
    \textfont2=\ninesy      \scriptfont2=\sixsy
                            \scriptscriptfont2=\fivesy
    \textfont3=\tenex       \scriptfont3=\tenex
                            \scriptscriptfont3=\tenex
    \textfont\itfam=\nineit \def\it{\fam\itfam\nineit}%
    \textfont\slfam=\ninesl \def\sl{\fam\slfam\ninesl}%
    \textfont\bffam=\ninebf \scriptfont\bffam=\sixbf
                            \scriptscriptfont\bffam=\fivebf
                            \def\bf{\fam\bffam\ninebf}%
    \textfont\ssfam=\niness \scriptfont\ssfam=\sixss
                            \scriptscriptfont\ssfam=\sixss
                            \def\ss{\fam\ssfam\niness}%
    \normalbaselineskip=12pt
    \setbox\strutbox=\hbox{\vrule height8.0pt depth3.0pt width0pt}%
    \let\big=\ninebig
    \normalbaselines\rm}

\def\eightpoint{\def\rm{\fam0\eightrm}%
    \textfont0=\eightrm      \scriptfont0=\sixrm
                             \scriptscriptfont0=\fiverm
    \textfont1=\eighti       \scriptfont1=\sixi
                             \scriptscriptfont1=\fivei
    \textfont2=\eightsy      \scriptfont2=\sixsy
                             \scriptscriptfont2=\fivesy
    \textfont3=\tenex        \scriptfont3=\tenex
                             \scriptscriptfont3=\tenex
    \textfont\itfam=\eightit \def\it{\fam\itfam\eightit}%
    \textfont\slfam=\eightsl \def\sl{\fam\slfam\eightsl}%
    \textfont\bffam=\eightbf \scriptfont\bffam=\sixbf
                             \scriptscriptfont\bffam=\fivebf
                             \def\bf{\fam\bffam\eightbf}%
    \textfont\ssfam=\eightss \scriptfont\ssfam=\sixss
                             \scriptscriptfont\ssfam=\sixss
                             \def\ss{\fam\ssfam\eightss}%
    \normalbaselineskip=10pt
    \setbox\strutbox=\hbox{\vrule height7.0pt depth2.0pt width0pt}%
    \let\big=\eightbig
    \normalbaselines\rm}

\def\tenbig#1{{\hbox{$\left#1\vbox to8.5pt{}\right.\n@space$}}}
\def\ninebig#1{{\hbox{$\textfont0=\tenrm\textfont2=\tensy
                       \left#1\vbox to7.25pt{}\right.\n@space$}}}
\def\eightbig#1{{\hbox{$\textfont0=\ninerm\textfont2=\ninesy
                       \left#1\vbox to6.5pt{}\right.\n@space$}}}

\font\sectionfont=cmbx10
\font\subsectionfont=cmti10

\def\figurecaptionfont{\ninepoint}
\def\tablecaptionfont{\ninepoint}
\def\footnotefont{\eightpoint}


\newcount\equationno
\newcount\bibitemno
\newcount\figureno
\newcount\tableno

\equationno=0
\bibitemno=0
\figureno=0
\tableno=0
\advance\pageno by -1


\footline={\ifnum\pageno=0{\hfil}\else
{\hss\rm\the\pageno\hss}\fi}


\def\section #1. #2 \par
{\vskip0pt plus .20\vsize\penalty-150 \vskip0pt plus-.20\vsize
\vskip 1.6 true cm plus 0.2 true cm minus 0.2 true cm
\global\def\equationlabel{#1}
\global\equationno=0
\leftline{\sectionfont #1. #2}\par
\immediate\write\terminal{Section #1. #2}
\vskip 0.7 true cm plus 0.1 true cm minus 0.1 true cm
\noindent}


\def\subsection #1 \par
{\vskip0pt plus 0.8 true cm\penalty-50 \vskip0pt plus-0.8 true cm
\vskip2.5ex plus 0.1ex minus 0.1ex
\leftline{\subsectionfont #1}\par
\immediate\write\terminal{Subsection #1}
\vskip1.0ex plus 0.1ex minus 0.1ex
\noindent}


\def\appendix #1 \par
{\vskip0pt plus .20\vsize\penalty-150 \vskip0pt plus-.20\vsize
\vskip 1.6 true cm plus 0.2 true cm minus 0.2 true cm
\global\def\equationlabel{\hbox{\rm#1}}
\global\equationno=0
\leftline{\sectionfont Appendix #1}\par
\immediate\write\terminal{Appendix #1}
\vskip 0.7 true cm plus 0.1 true cm minus 0.1 true cm
\noindent}


\def\enum{\global\advance\equationno by 1
(\equationlabel.\the\equationno)}


\def\ifundefined#1{\expandafter\ifx\csname#1\endcsname\relax}

\def\ref#1{\ifundefined{#1}?\immediate\write\terminal{unknown reference
on page \the\pageno}\else\csname#1\endcsname\fi}

\newwrite\terminal
\newwrite\bibitemlist

\def\bibitem#1#2\par{\global\advance\bibitemno by 1
\immediate\write\bibitemlist{\string\def
\expandafter\string\csname#1\endcsname
{\the\bibitemno}}
\item{[\the\bibitemno]}#2\par}

\def\beginbibliography{
\vskip0pt plus .15\vsize\penalty-150 \vskip0pt plus-.15\vsize
\vskip 1.2 true cm plus 0.2 true cm minus 0.2 true cm
\leftline{\sectionfont References}\par
\immediate\write\terminal{References}
\immediate\openout\bibitemlist=biblist
\frenchspacing\parindent=1.5em
\vskip 0.5 true cm plus 0.1 true cm minus 0.1 true cm}

\def\endbibliography{
\immediate\closeout\bibitemlist
\nonfrenchspacing\parindent=1.0em}


\def\figurecaption#1{\global\advance\figureno by 1
\narrower\figurecaptionfont
Fig.~\the\figureno.~#1}

\def\tablecaption#1{\global\advance\tableno by 1
\vbox to 0.5 true cm { }
\centerline{\tablecaptionfont%
Table~\the\tableno.~#1}
\vskip-0.4 true cm}

\def\thicktablerule{\hrule height1pt}
\def\thintablerule{\hrule height0.4pt}

\tenpoint

\immediate\openin\bibitemlist=biblist
\ifeof\bibitemlist\immediate\closein\bibitemlist
\else\immediate\closein\bibitemlist
\input biblist \fi


\def\thismonth{\ifcase\month\or
January\or February\or March\or April\or May\or June\or
July\or August\or September\or October\or November\or December\fi}



\def\rmd{{\rm d}}

\def\rme{{\rm e}}


\def\Romannumeral#1{\uppercase\expandafter{\romannumeral#1}}


\def\proof{\noindent{\sl Proof:}\kern0.6em}

\def\frac#1#2{\hbox{$#1\over#2$}}
\def\dual{\mathstrut^*\kern-0.1em}

\def\lvec#1{\setbox0=\hbox{$#1$}
    \setbox1=\hbox{$\scriptstyle\leftarrow$}
    #1\kern-\wd0\smash{
    \raise\ht0\hbox{$\raise1pt\hbox{$\scriptstyle\leftarrow$}$}}
    \kern-\wd1\kern\wd0}
\def\rvec#1{\setbox0=\hbox{$#1$}
    \setbox1=\hbox{$\scriptstyle\rightarrow$}
    #1\kern-\wd0\smash{
    \raise\ht0\hbox{$\raise1pt\hbox{$\scriptstyle\rightarrow$}$}}
    \kern-\wd1\kern\wd0}


\def\nab#1{{\nabla_{#1}}}
\def\nabstar#1{\nabla\kern-0.5pt\smash{\raise 4.5pt\hbox{$\ast$}}
               \kern-4.5pt_{#1}}

\def\drvstar#1{\partial\kern-0.5pt\smash{\raise 4.5pt\hbox{$\ast$}}
               \kern-5.0pt_{#1}}





\def\dirac#1{\gamma_{#1}}
\def\diracstar#1#2{
    \setbox0=\hbox{$\gamma$}\setbox1=\hbox{$\gamma_{#1}$}
    \gamma_{#1}\kern-\wd1\kern\wd0
    \smash{\raise4.5pt\hbox{$\scriptstyle#2$}}}


%
\rightline{CERN-TH/98-250}
\rightline{DESY 98-094}

\vskip 2.0 true cm minus 0.3 true cm
\centerline
{\bigbf Locality properties of Neuberger's lattice Dirac operator}
\vskip 1.5 true cm
\centerline{\bigrm Pilar Hern\'andez and Karl Jansen%
\footnote{$\raise5pt\hbox{$\scriptstyle\ast$}$}%
{\footnotefont Heisenberg foundation fellow}}
\vskip1ex
\centerline{\it CERN, Theory Division, CH-1211 Geneva, Switzerland}
\vskip 1.2 true cm
\centerline{\bigrm Martin L\"uscher}
\vskip1ex
\centerline{\it Deutsches Elektronen-Synchrotron DESY}
\centerline{\it Notkestrasse 85, D-22603 Hamburg, Germany}
\vskip 2.0 true cm
\centerline{\bf Abstract}
\vskip 1.5ex
The gauge covariant lattice Dirac operator $D$ 
which has recently been proposed
by Neuberger satisfies the Ginsparg-Wilson relation and
thus preserves chiral symmetry. The operator also avoids
a doubling of fermion species,
but its locality properties are not obvious.
We now prove that $D$ is local (with exponentially decaying tails)
if the gauge field is sufficiently smooth at the scale of the cutoff.
Further analytic and numerical studies moreover suggest
that the locality of the operator is in fact 
guaranteed under far more general conditions.
\vfill
\vskip 0.5 true cm
\eject

\section 1. Introduction

Many technical complications in the standard formulation of lattice QCD 
have to do with the fact that chiral symmetry is violated at the
scale of the cutoff. In particular, the quark masses are not
protected from additive renormalizations and the leading 
lattice effects in physical amplitudes are proportional to
the lattice spacing $a$ rather than being of order $a^2$.

Somewhat surprisingly 
it has recently turned out [\ref{HasenfratzI}--\ref{LuscherI}]
that chiral symmetry can be preserved on the lattice,
without fermion doubling, if the lattice Dirac operator $D$
satisfies a certain algebraic relation, 
$$
  \dirac{5}D+D\dirac{5}=aD\dirac{5}D,
  \eqno\enum
$$
originally due to Ginsparg and Wilson [\ref{GinspargWilson}].
We shall not discuss the significance and consequences
of this identity here, but
refer the reader to the original papers quoted above and 
the rapidly growing literature on the subject
[\ref{NarayananI}--\ref{Horvath}].
A point which should be emphasized however is that 
the Ginsparg-Wilson relation
only guarantees that the lattice theory 
has the same chiral symmetries as the continuum theory.
Locality, the correct behaviour in the classical continuum limit
and the absence of doubler modes 
are additional constraints which 
any decent lattice Dirac operator should satisfy.

Starting from the overlap formulation of chiral gauge theories,
a relatively simple
solution of the Ginsparg-Wilson relation 
has been found by Neuberger some time ago
[\ref{NeubergerI}].
Explicitly it is given by
$$
  D={1\over a}\bigl\{1-A(A^{\dagger}A)^{-1/2}\bigr\},
  \qquad
  A=1+s-aD_{\rm w},
  \eqno\enum
$$
where $D_{\rm w}$ denotes the standard Wilson-Dirac operator,
$$
  D_{\rm w}=\frac{1}{2}\left\{\dirac{\mu}(\nabstar{\mu}+\nab{\mu})
  -a\nabstar{\mu}\nab{\mu}\right\},
  \eqno\enum
$$
and $s$ is a real parameter in the range $|s|<1$
which will be fixed later
(cf.~appendix~A for unexplained notations).
Neuberger's operator is manifestly gauge covariant and 
can be shown to have no doubler modes. One can also easily verify
that it converges to the expected expression in the classical continuum limit,
up to a finite normalization constant, but
the requirement of locality is not obviously fulfilled.
Evidently it is very important to check that the operator is local
because the universality of the continuum limit depends on this 
fundamental property.

Before we begin with the detailed discussion of Neuberger's operator
it may be helpful to state what precisely is meant 
if we say that $D$ is local. Strict locality would imply
that the non-zero contributions to the sum
$$
  D\psi(x)=a^4\sum_{y}D(x,y)\psi(y)
  \eqno\enum
$$
come from the points $y$ in a finite neighbourhood of $x$.
Moreover the kernel $D(x,y)$ should only depend on the gauge field
variables residing near $x$.
From eqs.~(1.2) and (1.3) it is obvious, however, that
Neuberger's operator is not local in this restricted sense.
A more general definition of locality is hence adopted here, 
where the kernel is allowed to 
have exponentially decaying tails at large distances.
As long as the rate of decay can be shown to be proportional 
to the cutoff $1/a$, 
the sum in eq.~(1.4) will be completely dominated by the contributions
from a bounded region around $x$ with a fixed diameter in lattice units.
In particular, from the point of view of the continuum limit
there is little doubt that this kind of locality is as good as 
strict locality.

It is clear that a non-locality of the Neuberger operator
can only arise from the inverse square root of $A^{\dagger}A$ in eq.~(1.2).
Most of the time we shall thus be concerned with 
the properties of this operator.
In section~2 we first consider the case where $A^{\dagger}A$
is bounded from below by a positive constant.
Expanding the inverse square root of $A^{\dagger}A$ 
in a series of Legendre polynomials, 
the locality of $D$ may then be proved straightforwardly.
Moreover, the required lower bound on the spectrum of $A^{\dagger}A$ 
can be established rigorously
if the gauge field is sufficiently smooth at the scale of the cutoff
and Neuberger's operator is hence guaranteed to be local 
for all these fields.

One might expect that $D$ becomes 
increasingly non-local
when $A^{\dagger}A$ develops a zero mode.
We briefly examine this question in section~2, using
series expansions, and find that this is actually not so in general.
The numerical studies reported in section~3 confirm this
and they also provide a realistic estimate of the 
localization range of the operator at the gauge couplings of 
interest. 
All these results
fit into a simple picture which suggests 
that Neuberger's operator 
(with an appropriate choice of the parameter $s$)
is local for all statistically relevant
gauge field configurations.

\section 2. Spectrum of $A^{\dagger}A$ and locality of $D$

In this section some rigorous results are established
which show that $D$ is local 
(in the sense explained above)
under certain conditions. 
Along the way we shall find that 
the locality of $D$ is closely related to the spectral properties 
of $A^{\dagger}A$. This leads to interesting qualitative insights
which allow us to be more confident about the interpretation
of our numerical studies.

\subsection 2.1 Series expansion of $(A^{\dagger}A)^{-1/2}$

Following a suggestion of Bunk~[\ref{Bunk}] 
the inverse square root of $A^{\dagger}A$
may be expanded in a series of Legendre polynomials.
The expansion is suitable for numerical application,
but here we use it as a theoretical tool to discuss
the locality properties of the Neuberger operator.
In the following lines the detailed form of $A^{\dagger}A$
does not matter. To ensure the convergence of the
Legendre expansion we however assume that the bounds
\footnote{$\dagger$}{\footnotefont
Here and below 
an inequality between operators stands for the corresponding
inequality between the expectation values of the operators
in arbitrary normalizable states}
$$
  u\leq A^{\dagger}A\leq v
  \eqno\enum
$$
hold for some strictly positive constants $u<v$.
Whether this is the case for a given gauge field 
is a separate issue which will be addressed later.

The Legendre polynomials $P_k(z)$ may be defined through
the generating function
$$
  (1-2tz+t^2)^{-1/2}=\sum_{k=0}^{\infty}t^kP_k(z).
  \eqno\enum
$$
Usually $z$ is taken to be a number, but
eq.~(2.2) remains meaningful if we substitute
$$
  z=(v+u-2A^{\dagger}A)/(v-u).
  \eqno\enum
$$
It is easy to show that this operator has 
norm less than or equal to $1$.
In particular, using the operator calculus and 
the well-known properties
of the Legendre polynomials (as quoted in ref.~[\ref{GR}], for example),
this implies
$$
  \left\|P_k(z)\right\|\leq1.
  \eqno\enum
$$
The expansion (2.2)
is hence norm convergent for all $t$ satisfying $|t|<1$. 

We now introduce a parameter $\theta$ through
$$
  \cosh\theta=(v+u)/(v-u),
  \qquad\theta>0,
  \eqno\enum
$$
and set $t=\rme^{-\theta}$.
Eq.~(2.2) then assumes the form
$$
  (A^{\dagger}A)^{-1/2}=\kappa\,\sum_{k=0}^{\infty}t^kP_k(z),
  \qquad 
  \kappa=\left\{4t/(v-u)\right\}^{1/2},
  \eqno\enum
$$
since $1-2tz+t^2$ is proportional to $A^{\dagger}A$
for this choice of $t$.

\subsection 2.2 Legendre expansion and locality of $D$

We now show that the convergence of the 
Legendre expansion (2.6) and the strict locality of $A$ imply
that $D$ is local (with exponentially decaying tails).
The lattice is here taken to be infinitely extended in all directions.
In view of eq.~(1.2) it suffices to establish the locality 
of the inverse square root of $A^{\dagger}A$.

The kernel $G(x,y)$ which is associated with this operator,
$$
  (A^{\dagger}A)^{-1/2}\psi(x)=
  a^4\sum_yG(x,y)\psi(y),
  \eqno\enum
$$
is a matrix acting on the Dirac and colour indices
of the fermion field at the point $y$.
If we define the kernels $G_k(x,y)$ 
representing the operators $P_k(z)$ in the same way,
we have
$$
  G(x,y)=\kappa\,\sum_{k=0}^{\infty}t^kG_k(x,y).
  \eqno\enum
$$
It is easy to show that the norm convergence of the 
Legendre expansion implies the absolute convergence
of this series for all points $x$ and $y$.
Actually, from eq.~(2.4) one infers that 
$$
  a^4\left\|G_k(x,y)\right\|\leq1
  \quad\hbox{for all}\quad k,x,y,
  \eqno\enum
$$
where the norm here is the matrix norm in Dirac and colour space.

We now note that $G_k(x,y)$ vanishes unless $x$ and $y$ are sufficiently 
close to each other, because $P_k(z)$ is a polynomial in 
$A^{\dagger}A$ and $A$ is a combination of nearest-neighbour difference
operators. If we introduce the ``taxi driver distance"
$$
  \left\|x-y\right\|_1=\sum_{\mu}|x_{\mu}-y_{\mu}|,
  \eqno\enum
$$
the precise statement is that
$$
  G_k(x,y)=0\quad\hbox{for all}\quad k<\left\|x-y\right\|_1/2a.
  \eqno\enum
$$
Restricting the sum (2.8) to the non-zero terms
and recalling eq.~(2.9), the bound 
$$
  a^4\left\|G(x,y)\right\|\leq {\kappa\over1-t}
  \exp\left\{-\theta\left\|x-y\right\|_1/2a\right\}
  \eqno\enum
$$
is thus obtained. In particular,
the kernel is exponentially decaying at large distances
with a rate proportional to the cutoff $1/a$. Moreover its
dependence on the gauge field is local in a similar way, i.e.~up to 
exponentially small tails.

A technical detail we wish to emphasize is that 
the localization range $2a/\theta$ and the proportionality
constant in eq.~(2.12)
only depend on the bounds $u$ and $v$. As long as these can be
chosen uniformly in the gauge field, the 
Neuberger operator is guaranteed to behave 
essentially as a strictly local operator.
The differentiability
of the operator with respect to the gauge field 
can also be proved under these conditions
(appendix B).

\subsection 2.3 Bounds on $A^{\dagger}A$

The proof of the locality of $D$ given above depends on the 
convergence of the Legendre expansion and thus on the validity
of the bounds (2.1) for some positive constants $u<v$.
The upper bound is easily seen to hold for any gauge field
if we rewrite $A$ in the form
$$
  A=1+s+
  \sum_{\mu}\,\left\{\frac{1}{2}(1-\dirac{\mu})a\nab{\mu}-
  \frac{1}{2}(1+\dirac{\mu})a\nabstar{\mu}\right\}.
  \eqno\enum
$$
Using the triangle inequality it then follows that 
$$
  \|A\|=\|A^{\dagger}\|\leq8
  \eqno\enum
$$
for $|s|<1$
and $A^{\dagger}A$ is hence uniformly bounded from above.

As far as the lower bound is concerned it is clear that 
$A^{\dagger}A$ is non-negative, but one knows that
the operator can have zero modes for some gauge field configurations.
A uniform lower bound is hence excluded.
The strict positivity of $A^{\dagger}A$ may however
be established if the gauge field is sufficiently smooth at 
the scale of the cutoff. To make this more precise, let us suppose that 
$$
  \left\|1-U(p)\right\|\leq\epsilon 
  \quad\hbox{for all plaquettes $p$,}
  \eqno\enum
$$
where $U(p)$ denotes the product of the gauge field variables
around $p$ and the norm is the matrix norm in colour space.
For $s=0$ the inequality 
$$
  A^{\dagger}A\geq 1-30\epsilon
  \eqno\enum
$$
may then be proved (appendix C) and 
a similar bound may be deduced from this for general $s$
by substituting $A=s+A|_{s=0}$ and using triangle inequalities.
$A^{\dagger}A$ is hence uniformly bounded from below
by a positive constant if $\epsilon$ is sufficiently small.
In particular, the locality of $D$ is guaranteed under these conditions.

\subsection 2.4 Locality of $D$ in the presence of
                near-zero modes of $A^{\dagger}A$

When the gauge field strength is not small it can happen 
that some of the eigenvalues of $A^{\dagger}A$ 
are very close to zero or even equal to zero.
Since the exponent $\theta$ defined in subsect.~2.1
is proportional to $u^{1/2}$ at small $u$,
one is tempted to conclude that the locality of $D$ is lost 
in this situation.
We now show that this is not so in general.

Let us consider the case where 
the spectrum of $A^{\dagger}A$ is contained an 
interval $[u,v]$ except for an isolated eigenvalue 
$\lambda$ in the range
$$
  0<\lambda<\frac{1}{2}u.
  \eqno\enum
$$
As before the constants $u<v$ are some fixed positive numbers
while $\lambda$ is allowed to become arbitrarily small.
The projector on the associated eigenspace 
is given by
$$
  P=\oint{\rmd w\over2\pi i}\,(w-A^{\dagger}A)^{-1},
  \eqno\enum
$$
where the integration contour is a circle in the complex plane
centred at the origin with radius
$\frac{3}{4}u$. We now first prove that $P$ is local by noting 
that the operator in the square bracket
on the right-hand side of the identity
$$
  (w-A^{\dagger}A)^{-1}=(w^{\ast}-A^{\dagger}A)
  \left[(w^{\ast}-A^{\dagger}A)(w-A^{\dagger}A)\right]^{-1}
  \eqno\enum
$$
has eigenvalues between $(\frac{1}{4}u)^2$
and $(v+\frac{3}{4}u)^2$.
Its inverse may thus be expanded in a rapidly convergent series of 
Chebyshev polynomials and it follows
from this that the kernel associated with $P$ is local
with exponentially decaying tails. It should be emphasized that this
remains true even if $\lambda$ approaches zero, which is a regular
case in the above equations. In particular, the localization range of $P$
is determined by $u$ and $v$ alone.

To establish the  locality of $D$ we write
$$
  (A^{\dagger}A)^{-1/2}=
  (A^{\dagger}A)^{-1/2}P+(A^{\dagger}A)^{-1/2}(1-P)
  \eqno\enum
$$
and expand the terms on the right-hand
side of this equation in Legendre polynomials. 
In particular, 
$$
  (A^{\dagger}A)^{-1/2}(1-P)=\kappa\,\sum_{k=0}^{\infty}
  t^kP_k(z)(1-P),
  \eqno\enum
$$
where $t$, $z$ and $\kappa$ are as given in subsect.~2.1. 
The convergence 
of the series is guaranteed since $\|P_k(z)(1-P)\|\leq1$.
Proceeding essentially as in subsect.~2.2, 
this implies the locality of the operator (2.21).
For the Legendre expansion of the first term in eq.~(2.20),
the spectral bounds $u,v$ should be replaced by
$\tilde{u},\tilde{v}$, where
$$
  \tilde{v}=2\tilde{u},
  \qquad
  \tilde{u}<\lambda<\tilde{v}.
  \eqno\enum
$$
The associated expansion parameter $t$ is independent of $\tilde{u}$
and the localization range of the operator is hence the same for all
values of $\lambda$. Note that the divergence of the
normalization factor $\kappa$
cancels when the inverse square root of $A^{\dagger}A$
is multiplied with $A$. In particular, Neuberger's operator 
remains finite and local in the limit $\lambda\to0$.

\subsection 2.5 Summary

Perhaps the most important result obtained in this section concerns
the small field region where the bound (2.15) holds for some 
$\epsilon$ strictly less than $\frac{1}{30}$. 
Neuberger's operator is local in this case with exponentially decaying
tails. Moreover the localization range is uniformly bounded
from above by a constant depending on $\epsilon$ only.

In the large field region the situation appears to be more complicated
and it could be that the locality of $D$ cannot be guaranteed for 
all fields. 
Nevertheless we have been able to show that the presence of 
near-zero modes of $A^{\dagger}A$ does not by itself imply any non-locality.
Our analytical investigations rather suggest that $D$ remains local 
as long as the continuous spectrum 
of $A^{\dagger}A$ is separated from zero by a positive gap.

\section 3. Numerical studies of Neuberger's operator

In numerical simulations of quenched lattice QCD, using the
Wilson plaquette action, the representative gauge field
configurations at the gauge couplings of interest 
have relatively large average plaquette values.
As a consequence the locality of $D$ cannot be guaranteed
on the basis of the results of section~2 alone.
The purpose of the numerical studies reported here is 
to obtain some direct evidence for (or against) the locality of $D$
in this situation and to check whether the qualitative picture
is as suggested by the theoretical analysis.

The lattices that we have considered are of size $L$ in all directions
with periodic boundary conditions. $L/a$ has been set to $12$ or $16$.
The gauge group is taken to be SU(3), with three values 
of the bare coupling $g_0^2=6/\beta$ corresponding to $\beta=6.0,6.2$ 
and $6.4$. Following standard procedures, a representative ensemble
of statistically independent gauge field configurations 
has been generated for each lattice. It should be emphasized
that all results refer to quenched QCD. 
In the full theory the situation could be 
different, although there is currently no reason to expect this.

\subsection 3.1 Localization range of $D$

Let us consider the source field
$$
  \eta_{\alpha}(x)=\cases{1 & if $x=y$ and $\alpha=1$, \cr
                          0 & otherwise, \cr}
  \eqno\enum
$$
where $y$ is some particular point on the lattice and $\alpha$ 
runs over the colour and Dirac indices of the field.
We are then interested in the decay properties of 
$$
  \psi(x)=A(A^{\dagger}A)^{-1/2}\eta(x)
  \eqno\enum
$$
at large distances $\|x-y\|_1$ [cf.~eq.~(2.10)].
It is implicitly understood here
that the coordinate differences $x_{\mu}-y_{\mu}$ 
are taken modulo $L$ so as
to minimize the distance. In particular, the 
largest possible distance is $2L$. 

In fig.~1 we plot the expectation value of the function
$$
  f(r)=\max\left\{\|\psi(x)\|\bigm| \|x-y\|_1=r\right\}
  \eqno\enum
$$
for various values of $s$.
The norm $\|\psi(x)\|$ in this definition 
is the usual vector norm. To compute $\psi(x)$ we have used
a Chebyshev approximation for the inverse square root of $A^{\dagger}A$,
with coefficients adjusted so that 
a relative accuracy better than $10^{-9}$ 
is achieved [\ref{FoxParker},\ref{Recipes}].
One needs to know the extremal eigenvalues of $A^{\dagger}A$
for this, but as discussed below they can be calculated 
reliably with a modest effort. 
A technical point we wish to emphasize is
that the relatively high numerical precision
quoted above is required to avoid systematic effects in the 
calculated values of $f(r)$ at large distances.

\topinsert
\vbox{
\vskip-0.6true cm
 
\centerline{
\epsfxsize=12.0 true cm
\epsfbox{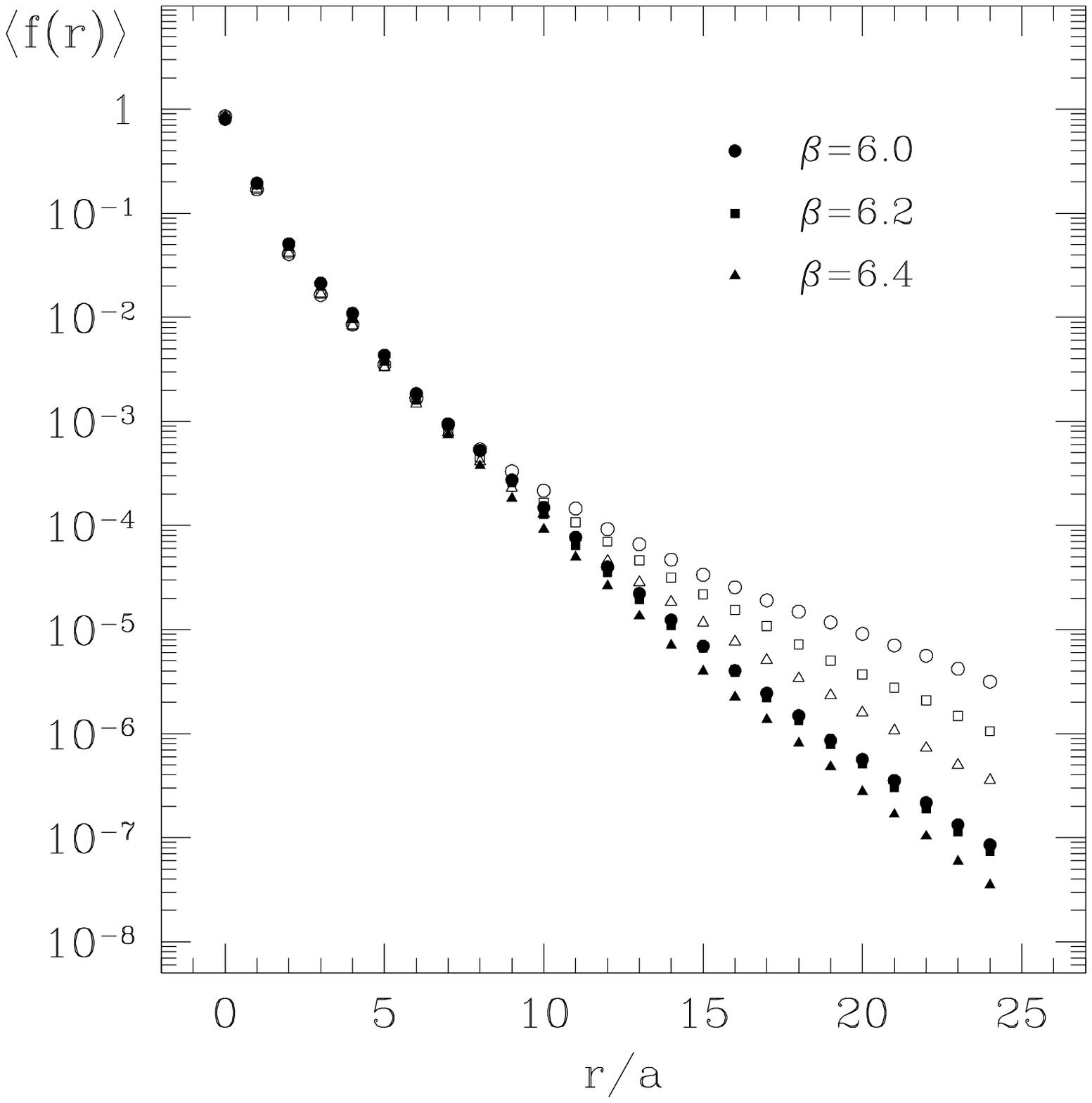}}
 
\vskip-1.0 true cm
\figurecaption{Expectation value of $f(r)$ [eq.~(3.3)] as a function of the 
distance $r$. Open symbols correspond to $s=0$ while the filled symbols
represent the data at $s=0.4$ ($\beta\leq6.2$) and $s=0.2$ 
($\beta=6.4$) respectively. The statistical errors are not visible 
on the scale of this plot.
}
}
\endinsert

In all cases considered $\langle f(r)\rangle$ is rapidly 
decaying when the distance $r$ increases.
Finite-volume effects appear to be negligible
here and significant differences between the curves
at different $\beta$ and $s$ are only seen
when $r/a$ is larger than $10$ or so.
For $r/a>13$ the data
can be represented by a single exponential,
$$
  \langle f(r)\rangle\propto \rme^{-\nu r/a},
  \eqno\enum
$$
with exponents $\nu$ as listed in table~1.
One may be worried at this point that 
the fluctuations of $f(r)$ are large, but our experience is 
that the mean deviations are at most a fraction of the average value
and would thus be barely visible in fig.~1.

From the figure and the table it is evident that $\langle f(r)\rangle$ 
becomes nearly independent of $\beta$ if $s$ is chosen
appropriately. The lowest curve in fig.~1 is also practically matched
by the curve that one obtains in the free quark theory.
In other words, as far as $\langle f(r)\rangle$ is concerned,
the localization properties of $D$ at $\beta\geq6.0$ 
and with a good choice of $s$ are roughly the same as in the free case.

\topinsert
\newdimen\digitwidth
\setbox0=\hbox{\rm 0}
\digitwidth=\wd0
\catcode`@=\active
\def@{\kern\digitwidth}
\vskip-2ex
\tablecaption{
Values of exponent $\nu$ [eq.~(3.4)]
}
\vskip1ex
$$\vbox{\settabs\+x&xxxxx&%
                  xxxxxx&xxxx&%
                  xxxxxx&xxxx&
                  xxxx&xxxxxxxxx&x\cr
\thicktablerule
\vskip1ex
                \+& \hfill     $\beta$ \hfill
                 && \hfill     $L/a$   \hfill
                 && \hfill     $s$     \hfill
                 && \hfill     $\nu$   \hfill
                 &  \cr
\vskip1.0ex
\thintablerule
\vskip1.5ex
  \+& \hfill $6.0$ \hfill
  &&  \hfill $12$  \hfill
  &&  \hfill $0.0$ \hfill
  &&  \hfill $0.28$ \hfill
  &\cr
  \+& \hfill $$ \hfill
  &&  \hfill $12$  \hfill
  &&  \hfill $0.4$ \hfill
  &&  \hfill $0.49$ \hfill
  &\cr
  \+& \hfill $$ \hfill
  &&  \hfill $12$  \hfill
  &&  \hfill $0.6$ \hfill
  &&  \hfill $0.45$ \hfill
  &\cr
\vskip0.5ex
  \+& \hfill $6.2$ \hfill
  &&  \hfill $12$  \hfill
  &&  \hfill $0.0$ \hfill
  &&  \hfill $0.35$ \hfill
  &\cr
  \+& \hfill $$ \hfill
  &&  \hfill $12$  \hfill
  &&  \hfill $0.4$ \hfill
  &&  \hfill $0.49$ \hfill
  &\cr
  \+& \hfill $$ \hfill
  &&  \hfill $12$  \hfill
  &&  \hfill $0.6$ \hfill
  &&  \hfill $0.42$ \hfill
  &\cr
\vskip0.5ex
  \+& \hfill $6.4$ \hfill
  &&  \hfill $16$  \hfill
  &&  \hfill $0.0$ \hfill
  &&  \hfill $0.40$ \hfill
  &\cr
  \+& \hfill $$ \hfill
  &&  \hfill $12$  \hfill
  &&  \hfill $0.0$ \hfill
  &&  \hfill $0.40$ \hfill
  &\cr
  \+& \hfill $$ \hfill
  &&  \hfill $12$  \hfill
  &&  \hfill $0.2$ \hfill
  &&  \hfill $0.53$ \hfill
  &\cr
  \+& \hfill $$ \hfill
  &&  \hfill $12$  \hfill
  &&  \hfill $0.4$ \hfill
  &&  \hfill $0.49$ \hfill
  &\cr
\vskip1ex
\thicktablerule
}$$
\endinsert

That some tuning of $s$ is required to
preserve the localization range of $D$ 
does not come as a total surprise,
because the spectrum of the Wilson-Dirac operator $D_{\rm w}$
moves to the right in the complex plane when $\beta$ is decreased.
In particular, 
the critical bare mass $m_{\rm c}$ is shifted
to $-0.68/a$, $-0.74/a$ and $-0.82/a$ at $\beta=6.4$, $6.2$ and 
$6.0$ respectively [\ref{AlltonEtAl}]. Since 
$$
  A=-a(D_{\rm w}+m_0),
  \qquad
  m_0=-(1+s)/a,
  \eqno\enum
$$
a positive value of $s$ partly compensates for this and ensures
that $m_0$ keeps away from $m_{\rm c}$ by an appreciable margin.

\subsection 3.2 Spectrum of $A^{\dagger}A$

To make contact with the theoretical discussion of section~2 
we now proceed to examine the distribution of 
the low-lying eigenvalues of $A^{\dagger}A$.
For any given gauge field configuration 
these eigenvalues can be computed
by minimizing the Ritz functional using a conjugate gradient algorithm.
The method has previously been applied and is described in detail in 
refs.~[\ref{BunkEtAl},\ref{KalkreuterSimma}]. 

\topinsert
\vbox{
\vskip-1.4 true cm
 
\centerline{
\epsfxsize=14.0 true cm
\epsfbox{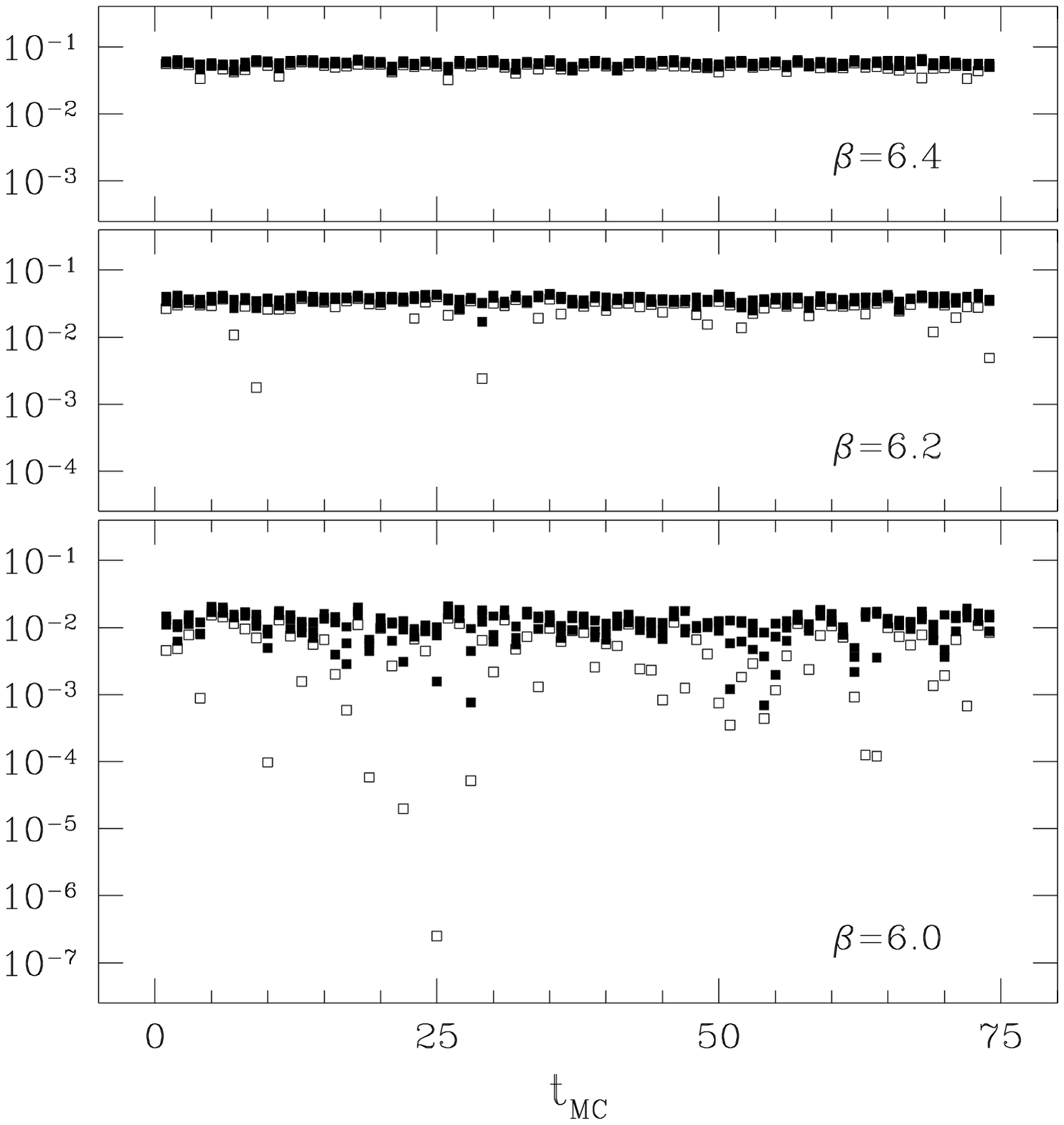}}
 
\vskip-1.0 true cm
\figurecaption{
Monte Carlo time evolution of the four lowest eigenvalues of $A^{\dagger}A$. 
The smallest eigenvalue is represented
by an open square and the higher ones by filled squares.
All data are from the $12^4$ lattice with $s=0$.
}
}
\endinsert

As shown by fig.~2 the spectrum of $A^{\dagger}A$ 
depends quite strongly on the gauge coupling. At $\beta=6.4$
the lower end of the spectrum appears to be clearly separated from zero
and $A^{\dagger}A$ thus satisfies the bounds~(2.1).
This is not so at $\beta=6.2$ and $6.0$, where
one has a non-zero probability to find eigenvalues that are orders of 
magnitude below the rest of the spectrum.
Moreover the band of the ordinary modes is 
wider at these couplings.

When $s$ is set to higher values (such as those quoted in fig.~1)
the qualitative features of the 
spectrum do not change, but the level of the ordinary low-lying eigenvalues 
is raised by a factor $2$ or so.
The maximal eigenvalue, on the other hand, hardly changes and 
stays around $41$ for all $\beta$ and $s$ that we have considered.
So far we have only been able to analyse a limited
number of gauge field configurations on the larger lattice
and thus cannot make a detailed statement about the 
volume dependence of the spectrum at this point.
There is, however, a clear tendency that the probability for
near-zero modes increases with the lattice size.

\subsection 3.3 Localization properties of the near-zero modes

The minimization of the Ritz functional not only 
yields the low-lying eigenvalues of $A^{\dagger}A$ 
but also the corresponding eigenfunctions.
In particular, the localization properties of the 
near-zero modes can be studied straightforwardly.
A possible definition of the localization range of 
a given wave function is discussed in ref.~[\ref{Jansen}].
Proceeding along these lines we have found 
that all near-zero modes are well localized 
with exponentially decaying tails.
Moreover we have observed that the 
localization ranges shrink significantly 
when $s$ is increased. All this completely agrees with the results 
previously reported by Edwards et al.~[\ref{EdwardsEtAl}] (see also ref.~[\ref{Jansen}]).

For illustration let us consider an eigenvector 
$\phi(x)$, suitably normalized, with maximal magnitude
$\|\phi(x)\|$ at some point $x=y$.
The function 
$$
  h(r)=\max\left\{\|\phi(x)\|\bigm| \|x-y\|_1=r\right\}
  \eqno\enum
$$
then provides an upper bound on the wave function
at distance~$r$ from the centre of its localization region.
A typical result for $h(r)$ is plotted in fig.~3. The
associated eigenvalue is nearly two orders of magnitude below the 
band of the ordinary low-lying modes in this example.

\topinsert
\vbox{
\vskip-0.6true cm plus 0.5 true cm
 
\centerline{
\epsfxsize=12.0 true cm
\epsfbox{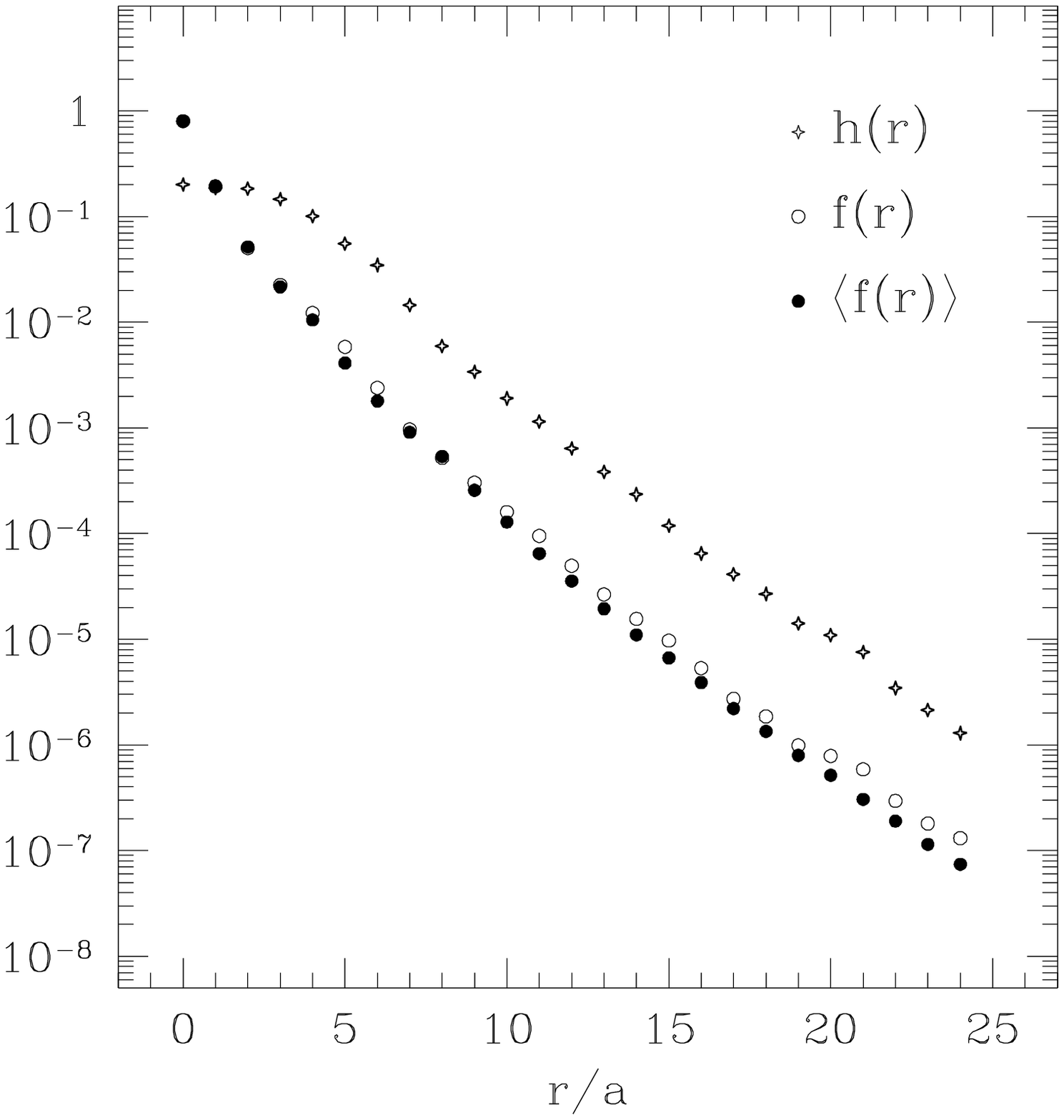}}
 
\vskip-1.2 true cm
\figurecaption{Plot of the magnitude $h(r)$ 
of a typical near-zero mode at 
$\beta=6.2$, $s=0.4$ and $L/a=12$,
together with the function $f(r)$ and the 
corresponding ensemble average~$\langle f(r)\rangle$.
}
}
\endinsert

In fig.~3 the function $f(r)$, calculated for the same gauge
field configuration and with the same choice of $y$,
is also shown. Comparing with $\langle f(r)\rangle$
one clearly sees that the presence of the near-zero mode does not affect 
the localization properties of $D$ in any significant way.

\subsection 3.4 Synthesis

Taken together the theoretical and numerical results reported in this paper
suggest that the spectrum of $A^{\dagger}A$ is clearly separated from zero
at large $\beta$, for all statistically relevant gauge field configurations.
The locality of Neuberger's operator is rigorously guaranteed
in this situation.
It is our experience, however, that the 
theoretical bounds tend to over-estimate the localization range 
by a large factor. At $\beta=6.4$ and $s=0$, for example, 
the exponent $\theta/2=0.03$ which one obtains 
by inserting the numerically determined 
spectral bounds in eq.~(2.5) is much smaller than the 
exponent $\nu=0.4$ quoted in table~1.

At the lower values of $\beta$, near-zero
modes develop and become increasingly frequent,
but they are well localized (if $s$ is chosen appropriately) and thus do
not destroy the locality of Neuberger's operator.
On the lattices that we have studied
the near-zero modes are isolated from the rest of the spectrum.
The observed locality properties are, therefore, completely
in line with the theoretical discussion of section~2, although
here again the analytical estimates of the localization radius 
are far too pessimistic. 

An important point to note in this context is
that the probability to find localized near-zero modes grows 
proportionally to the volume, because widely separated regions
on the lattice basically behave as separate systems.
Eventually there will be configurations with several near-zero modes
and in the infinite volume limit these modes form a dense 
so-called pure point spectrum 
(cf.~ref.~[\ref{ReedSimon}], chapter \Romannumeral{7}).
It seems unlikely that this statistical phenomenon
has any relevance for the locality properties of Neuberger's operator,
but it is clearly desirable to check this 
by extending our studies to larger lattices.

\section 4. Concluding remarks

At this point there is little doubt that Neuberger's operator
is local at small gauge couplings $g_0$,
although a rigorous proof of the locality is only available 
if the gauge field is assumed to satisfy the smoothness condition (2.15)
for some $\epsilon<\frac{1}{30}$.
This constraint can, incidentally, be imposed on the system 
by choosing an appropriate gauge field action
and one then obtains a lattice regularization of QCD 
which preserves the chiral and flavour symmetries without
violating basic principles. As far as we know
all other regularizations of QCD that have been considered to date
break chiral symmetry and the folklore has been that this is 
in fact unavoidable.

For numerical simulations of lattice QCD with the standard gauge action,
the locality of Neuberger's operator must be guaranteed for
larger values of the gauge coupling as well.
We have addressed this question in quenched QCD and did not
find any indication that the locality is lost at the couplings
of interest, provided the parameter $s$ is chosen appropriately.

In principle one may now use Neuberger's operator
to calculate the hadron spectrum etc., but this may still
be somewhat premature, because there are other choices of 
$A$ which lead to significantly smaller localization ranges,
at least in the free case [\ref{Bietenholz},\ref{Niedermayer}].
This may be important in practice, since one cannot afford to 
simulate very large lattices. Moreover, as has been pointed out
by Niedermayer [\ref{Niedermayer}], the probability for near-zero modes
may be very much suppressed for some of the proposed choices of $A$
and this too could make the calculations easier.

\vskip1ex
We are indebted to Peter Hasenfratz, Ferenc Niedermayer and Peter Weisz
for encouragement and many helpful discussions. M.~L.~would
like to thank the Institute for Theoretical Physics at the University
of Bern for hospitality in June, where part of this paper has been 
completed. The computer resources for this project have been provided
by CERN and CIEMAT. We thank these instutions for their support.

\appendix A

The notational conventions used in this paper are standard. 
We consider a four-dimensional hyper-cubic lattice with spacing $a$
and variable size. If the lattice is finite we impose periodic
boundary conditions although most results hold for
other boundary conditions as well.
The gauge field is represented
by unitary matrices $U(x,\mu)$ where $x$ runs through all lattice points
and $\mu=0,\ldots,3$ labels the space-time directions.
Dirac fields $\psi(x)$ carry a Dirac and a colour index
as in the continuum theory. 
The gauge covariant forward and backward difference operators 
act on such fields according to
$$
  \eqalignno{
  \nab{\mu}\psi(x)&=
  {1\over a}\bigl[U(x,\mu)\psi(x+a\hat{\mu})-\psi(x)\bigr],
  &\enum\cr
  \noalign{\vskip2ex}
  \nabstar{\mu}\psi(x)&=
  {1\over a}\bigl[\psi(x)-U(x-a\hat{\mu},\mu)^{-1}
  \psi(x-a\hat{\mu})\bigr],
  &\enum\cr}
$$
where $\hat{\mu}$ denotes the unit vector in direction $\mu$.
Since we are in euclidean space,
the Dirac matrices can be taken to be hermitean,
$$
  \diracstar{\mu}{\dagger}=\dirac{\mu},
  \qquad
  \{\dirac{\mu},\dirac{\nu}\}=2\delta_{\mu\nu},
  \eqno\enum
$$
and our conventions for $\dirac{5}$ and $\sigma_{\mu\nu}$ are
$$
  \dirac{5}=\dirac{0}\dirac{1}\dirac{2}\dirac{3},
  \qquad
  \sigma_{\mu\nu}={i\over2}\left[\dirac{\mu},\dirac{\nu}\right].  
  \eqno\enum
$$
Repeated indices are always summed over unless stated otherwise.

\appendix B

The expansion in Legendre polynomials derived in sect.~2
is exponentially convergent. It is not obvious, however,
that it may be differentiated with respect to the gauge field
(or any other parameter), because the differentiated 
polynomials need not be uniformly bounded.
In the following lines we establish a bound 
which excludes such an irregular behaviour.

We first derive an integral representation 
for the Legendre polynomials.
Starting from eqs.~(2.2) and (2.3) we have
$$
  P_k(z)=\oint{\rmd w\over 2\pi i}\, w^{-k-1}
  \left(w^2-2wz+1\right)^{-1/2},
  \eqno\enum
$$
where the integration runs along a circle in the complex plane
centred at the origin. The radius $r$ of the circle should be strictly
less than $1$ to avoid the singularities of the integrand.
Because of the square root and since $z$ is an operator, 
eq.~(B.1) is not easily differentiated.
To overcome this difficulty we make use of a well-known identity
to rewrite the integral in the form 
$$
  P_k(z)=\int_{-\infty}^{\infty}{\rmd\sigma\over\pi}\,
  \oint{\rmd w\over 2\pi i}\, w^{-k-1}
  (w^2-2wz+1+\sigma^2)^{-1}.
  \eqno\enum
$$
Note that the denominator of the integrand 
can be factorized according to 
$$
  w^2-2wz+1+\sigma^2=(w-u^{\dagger})(w-u),
  \qquad
  u=z+i\left(1+\sigma^2-z^2\right)^{1/2}.
  \eqno\enum
$$
Since $u^{\dagger}u=1+\sigma^2$ it is then immediately clear that 
$$
  \|(w^2-2wz+1+\sigma^2)^{-1}\|\leq 
  \bigl\{\left(1+\sigma^2\right)^{1/2}-r\bigr\}^{-2}.
  \eqno\enum
$$
In particular, the integral~(B.2) is norm convergent.

Let us now assume that $A^{\dagger}A$ depends on some parameter
$\tau$ in a differentiable manner such that 
$\dot{z}=\partial z/\partial\tau$ has finite norm.
An upper bound on the derivative of 
the Legendre polynomials with respect to $\tau$
is then obtained by differentiating eq.~(B.2) 
and applying eq.~(B.4). The right-hand side of the resulting inequality
$$
  \|\dot{P}_k(z)\|\leq 4\left\|\dot{z}\right\|r^{-k}
  \int_{0}^{\infty}{\rmd\sigma\over\pi}\,
  \bigl\{\left(1+\sigma^2\right)^{1/2}-r\bigr\}^{-4}
  \eqno\enum
$$
can be evaluated by substituting
$$
  (1+\sigma^2)^{1/2}-1=(1-r)\rho^2, 
  \qquad 0\leq\rho<\infty.
  \eqno\enum
$$
After some algebra one then ends up with the bound
$$
  \|\dot{P}_k(z)\|\leq \hbox{constant}\times\left\|\dot{z}\right\|r^{-k}
  (1-r)^{-4},
  \eqno\enum
$$
where the constant is independent of $k$ and $r$.

So far the radius $r$ has not been specified apart from the requirement
that it should be in the range $0<r<1$. We may now
adjust the radius so that the factor $r^{-k}(1-r)^{-4}$ is minimized.
Using simple estimates this leads to the bound
$$
  \|\dot{P}_k(z)\|\leq \hbox{constant}\times\left\|\dot{z}\right\|
  (1+k)^4.
  \eqno\enum
$$
The differentiated series~(2.6) is hence 
exponentially convergent with the same exponent as the 
original series. Similar estimations show that this is also true when 
higher-order differential operators are applied
(each differentiation gives rise to an additional factor of $(1+k)^2$ 
in the bound on the Legendre polynomials).

\appendix C

To establish eq.~(2.16) we first expand the product 
$A^{\dagger}A$ using simple identities for the 
covariant difference operators and the Dirac matrices.
As a result one gets a sum of terms,
$$
  A^{\dagger}A= 
  1+\frac{1}{4}\sum_{\mu\neq\nu}
  \left\{B_{\mu\nu}+C_{\mu\nu}+D_{\mu\nu}\right\},
  \eqno\enum
$$
which can be treated separately (setting $s=0$ has been essential here
to ensure the cancellation of some non-trivial diagonal terms).
Explicitly they are given by
$$
  \eqalignno{
  B_{\mu\nu}&=
  a^4\nabstar{\mu}\nab{\mu}\nabstar{\nu}\nab{\nu},
  &\enum\cr\noalign{\vskip2ex}
  C_{\mu\nu}&=\frac{1}{2}i\sigma_{\mu\nu}
  a^2\left[\nabstar{\mu}+\nab{\mu},\nabstar{\nu}+\nab{\nu}\right],
  &\enum\cr\noalign{\vskip2ex}
  D_{\mu\nu}&=-\dirac{\mu}
  a^2\left[\nabstar{\mu}+\nab{\mu},\nabstar{\nu}-\nab{\nu}\right].
  &\enum
  }
$$
The commutator terms (C.3) and (C.4) are proportional to
the field strength and should thus be of order $\epsilon$.
To prove this we note that
$$
  \eqalignno{
  &a^2\left[\nab{\mu},\nab{\nu}\right]\psi(x)=
  &\cr\noalign{\vskip2ex}
  &\quad\bigl\{U(x,\mu)U(x+a\hat{\mu},\nu)-U(x,\nu)U(x+a\hat{\nu},\mu)\bigr\}
  \psi(x+a\hat{\mu}+a\hat{\nu}).
  &\enum}
$$
The curly bracket in this equation 
is equal to a unitary matrix times $1-U(p)$ 
for some plaquette $p$. Eq.~(2.15) thus implies the bound
$$
  \left\|a^2\left[\nab{\mu},\nab{\nu}\right]\right\|\leq\epsilon
  \eqno\enum
$$
and the same inequality also holds if one or both 
forward difference operators are replaced by backward
difference operators. In particular,
$$
  \left\|C_{\mu\nu}\right\|\leq2\epsilon
  \quad\hbox{and}\quad
  \left\|D_{\mu\nu}\right\|\leq4\epsilon.
  \eqno\enum
$$
To bound the first term, eq.~(C.2), we rewrite it in the form
$$
  B_{\mu\nu}=a^4\nabstar{\mu}\nabstar{\nu}\nab{\nu}\nab{\mu}
  -a^3\nabstar{\mu}\left[\nab{\mu},\nabstar{\nu}-\nab{\nu}\right],
  \eqno\enum
$$
which shows that it is equal to a non-negative operator
plus another operator with norm less than $4\epsilon$. 
Taken together these estimates imply that the right-hand side
of eq.~(C.1) is bounded from below by $1-30\epsilon$
which proves eq.~(2.16).


\beginbibliography

\bibitem{GinspargWilson}
P. H. Ginsparg and K. G. Wilson,
Phys. Rev. D25 (1982) 2649

\bibitem{HasenfratzI}
P. Hasenfratz,
Nucl. Phys. B (Proc. Suppl.) 63A-C (1998) 53

\bibitem{HasenfratzEtAl}
P. Hasenfratz, V. Laliena and F. Niedermayer,
Phys. Lett. B427 (1998) 125

\bibitem{NeubergerI}
H. Neuberger,
Phys. Lett. B417 (1998) 141;
{\it ibid}\/ B427 (1998) 353

\bibitem{HasenfratzII}
P. Hasenfratz,
Nucl. Phys. B525 (1998) 401

\bibitem{LuscherI}
M. L\"uscher,
Phys. Lett. B428 (1998) 342

\bibitem{NarayananI}
R. Narayanan,
Phys. Rev. D58 (1998) 97501

\bibitem{Bietenholz}
W. Bietenholz,
Solutions of the Ginsparg-Wilson relation and improved domain wall fermions,
hep-lat/9803023

\bibitem{HasenfratzIII}
P. Hasenfratz,
The theoretical background and properties of perfect actions,
hep-lat/9803027

\bibitem{LangEtAl}
F. Farchioni, C. B. Lang, M. Wohlgenannt, 
Phys. Lett. B433 (1998) 377

\bibitem{Chiu}
T. W. Chiu,
Phys. Rev. D58 (1998) 74511

\bibitem{Chandrasekharan}
S. Chandrasekharan,
Lattice QCD with Ginsparg-Wilson fermions,
\hfill\break
hep-lat/9805015

\bibitem{Splittorff}
K. Splittorff and A. D. Jackson,
The Ginsparg-Wilson relation and local chiral random
matrix theory,
hep-lat/9806018

\bibitem{KikukawaYamada}
Y. Kikukawa and A. Yamada,
Weak coupling expansion of massless QCD with 
a Ginsparg-Wilson fermion and axial U(1) anomaly,
hep-lat/9806013

\bibitem{ChiuZenkin}
T. W. Chiu and S. V. Zenkin,
On solutions of the Ginsparg-Wilson relation,
hep-lat/9806019

\bibitem{NeubergerIII}
H. Neuberger,
Phys. Rev. Lett. 81 (1998) 4060

\bibitem{EdwardsEtAlII}
R. G. Edwards, U. M. Heller and R. Narayanan,
A study of practical implementations of the overlap Dirac operator 
in four dimensions,
hep-lat/9807017

\bibitem{Horvath}
I. Horvath,
Phys. Rev. Lett. 81 (1998) 4063               

\bibitem{Bunk}
B. Bunk,
Nucl. Phys. B (Proc. Suppl.) 63A-C (1998) 952

\bibitem{GR}
I. S. Gradshteyn and I. M. Ryshik,
Table of Integrals, Series and Products
(Academic Press, New York, 1965)
 
\bibitem{FoxParker}
L. Fox and I. B. Parker, 
Chebyshev polynomials in numerical analysis
(Oxford University Press, London, 1968)

\bibitem{Recipes}
W. H. Press, S. A. Teukolsky, W. T. Vetterling and B. P. Flannery,
Numerical recipes,
$2^{\rm nd}$ edition 
(Cambridge University Press, Cambridge, 1992)

\bibitem{AlltonEtAl}
C. R. Allton, V. Gimenez, L. Giusti and F. Rapuano,
Nucl. Phys. B489 (1997) 427

\bibitem{BunkEtAl}
B. Bunk, K. Jansen, M. L\"uscher and H. Simma, 
Conjugate gradient algorithm to compute the low-lying eigenvalues
of the Dirac operator in lattice QCD,
ALPHA collaboration internal report (1994),
unpublished

\bibitem{KalkreuterSimma}
T. Kalkreuter and H. Simma, Comp. Phys. Comm. 93 (1996) 33

\bibitem{EdwardsEtAl}
R. G. Edwards, U. M. Heller and R. Narayanan,
Nucl. Phys. B535 (1998) 403

\bibitem{Jansen}
K. Jansen, C. Liu, H. Simma and D. Smith, 
Nucl. Phys. B (Proc. Suppl.) 53 (1997) 262; H. Simma and D. Smith,
Low-lying Eigenvalues of the improved Wilson-Dirac Operator of QCD,
hep-lat/9801025

\bibitem{Niedermayer}
F. Niedermayer,
Exact chiral symmetry, topological charge and related topics,
plenary talk given at the International Symposium on Lattice Field Theory,
Boulder, July 13-18, 1998,
hep-lat/9810026

\bibitem{ReedSimon}
M. Reed and B. Simon, Methods of modern mathematical physics
(Academic Press, New York, 1972)

\endbibliography

\bye